\documentclass[pdflatex,sn-nature]{sn-jnl}

\usepackage{graphicx}%
\usepackage{multirow}%
\usepackage{amsmath,amssymb,amsfonts}%
\usepackage{amsthm}%
\usepackage{mathrsfs}%
\usepackage[title]{appendix}%
\usepackage{xcolor}%
\usepackage{textcomp}%
\usepackage{booktabs}%
\usepackage{algorithm}%
\usepackage{algorithmicx}%
\usepackage{algpseudocode}%
\usepackage{listings}%
\usepackage{braket} %

\begin{document}

\title[Article Title]{Unveiling Hidden Vulnerabilities in Quantum Systems by Expanding Attack Vectors through Heisenberg’s Uncertainty Principle}

\author*[1,2]{\fnm{Jose Roberto} \sur{Rosas Bustos}}\email{jrosasbu@uwaterloo.ca}

\author[1,2]{\fnm{Jesse} \spfx{Van} \sur{Griensven Thé}}\email{jesse.the@uwaterloo.ca}

\author[1]{\fnm{Roydon Andrew} \sur{Fraser}}\email{rfraser@uwaterloo.ca}

\affil*[1]{\orgname{University of Waterloo}, \orgaddress{\country{Canada}}}
\affil[2]{\orgname{LAKES Environmental Research Inc.}, \orgaddress{\country{Canada}}}

\abstract{This study uncovers novel vulnerabilities within Quantum Key Distribution (QKD) protocols that extend beyond traditional implementation flaws, such as loopholes. These newly identified vulnerabilities arise from the complex interaction between Bell Inequalities (BIs) and Hidden Variable Theories (HVTs), further exacerbated by the Heisenberg Uncertainty Principle (HUP). Through a combination of theoretical analysis, simulations, and quantum experiments, we reveal critical security weaknesses that challenge the core assumptions of today's quantum cryptography. While these vulnerabilities differ from known loopholes, when considered alongside them and traditional cyberattacks, they present a significant threat to the robustness of QKD and quantum integrity systems. These results provide a new perspective to rethink current quantum security frameworks to ensure the robustness of future quantum cryptographic and quantum integrity protocols.}

\keywords{Quantum Key Distribution, Bell Inequalities, Hidden Variable Theories, Quantum Cryptography, Heisenberg Uncertainty, Quantum Security, Attack Strategies, Quantum System Weaknesses, Quantum Integrity}

\maketitle
\section{Introduction}\label{intro}
\begin{figure}
    \centering
    \includegraphics[width=1\textwidth]{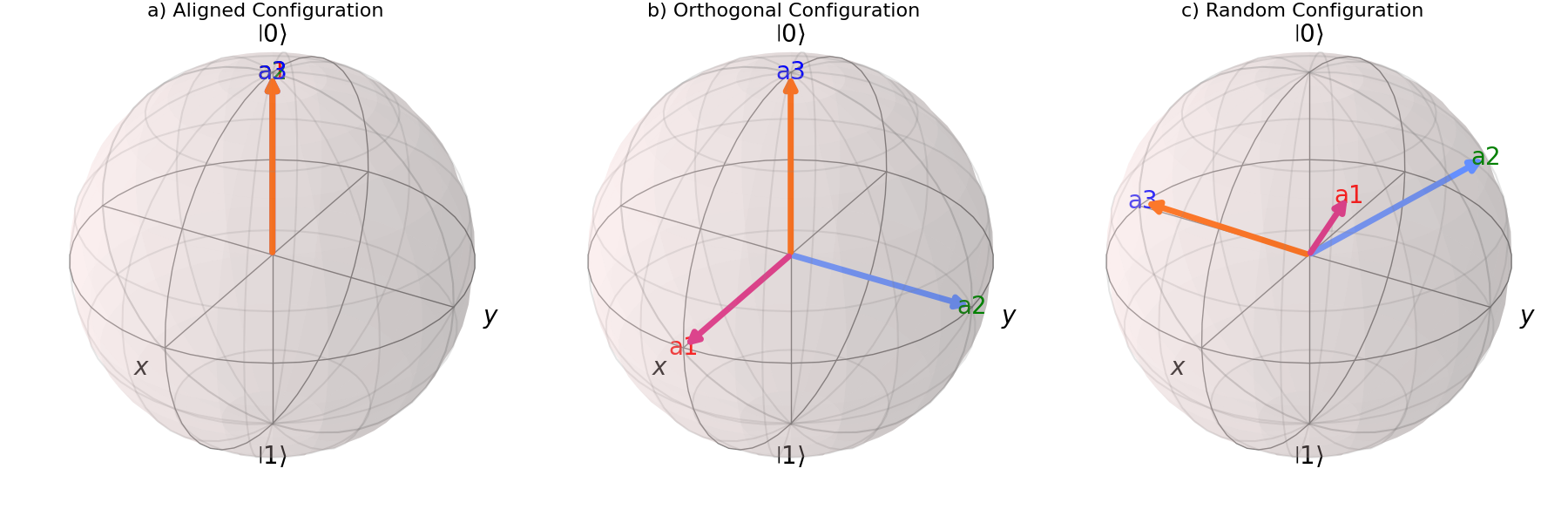}
    \caption{Visualization of the different measurement configurations on the Bloch sphere. (a) Aligned Configuration, where all measurement vectors are aligned along the same axis. (b) Orthogonal Configuration, where measurement vectors are orthogonal to each other. (c) Random Configuration, where measurement vectors are randomly oriented.}
    \label{fig:measurement_configurations}
\end{figure}

Quantum Key Distribution (QKD) \cite{scarani_security_2009}, a cornerstone of secure communication, leverages quantum mechanics principles, such as entanglement and the violation of Bell Inequalities (BIs), to detect eavesdropping and ensure security \cite{ekert_quantum_1991, aspect_experimental_1982}. However, the security framework of QKD protocols faces potential vulnerabilities due to the intersection of BIs with Hidden Variable Theories (HVTs) \cite{clauser_proposed_1969, brunner_bell_2014} and the implications of the Heisenberg Uncertainty Principle (HUP). As shown in Figure \ref{fig:measurement_configurations}, different measurement configurations can reveal overlaps between quantum mechanics and HVTs, particularly within vicinities defined by the HUP, which could lead to security vulnerabilities in quantum cryptographic protocols.

The HUP \cite{busch_colloquium_2014} imposes fundamental limits on the simultaneous measurement precision of certain pairs of properties in quantum systems, including entangled states, which, when combined with assumptions in Bell's theorem, particularly those arising from the Einstein-Podolsky-Rosen (EPR) paradox \cite{einstein_can_1935}, may expose weaknesses in quantum cryptographic systems. This convergence between BIs and HVTs, exacerbated by HUP, suggests that QKD and quantum integrity protocols may not be as secure as previously assumed, potentially allowing for new attack vectors.

\subsection{Research Question} 
This study investigates how the convergence of Bell Inequalities and Hidden Variable Theories, particularly in the context of the Heisenberg Uncertainty Principle, impacts the security of Quantum Key Distribution systems and Quantum Integrity.

\subsection{Hypothesis} 
It is hypothesized that specific convergence points between BIs and HVTs within quantum systems introduce exploitable vulnerabilities, further amplified by the Heisenberg Uncertainty Principle. 

\subsection{Objectives}
This study aims to achieve two primary objectives: 
\begin{enumerate}
    \item First, to reassess the theoretical vulnerabilities inherent in the assumptions underlying BIs and HVTs, particularly where their predictions converge \cite{reid_colloquium_2009}.
    \item Second, to extend this analysis by exploring the vicinities of these convergence points through the lens of HUP, thereby uncovering additional vulnerabilities and providing insights that could inform the development of more robust quantum cryptographic protocols.
\end{enumerate}

\section{Results}\label{results}
This section details our findings, revealing specific vulnerabilities in QKD protocols arising from the convergence of Bell Inequalities (BIs) and Hidden Variable Theories (HVTs) within multipartite discrete variable quantum systems. By examining cases where predictions from quantum mechanics align with those from HVTs, we identify critical weaknesses in the underlying security assumptions of these quantum cryptographic protocols.

\subsection{Convergence of BIs and HVTs in Multipartite Discrete Variable Quantum Systems}
\begin{figure}
    \centering
    \includegraphics[width=0.7\textwidth]{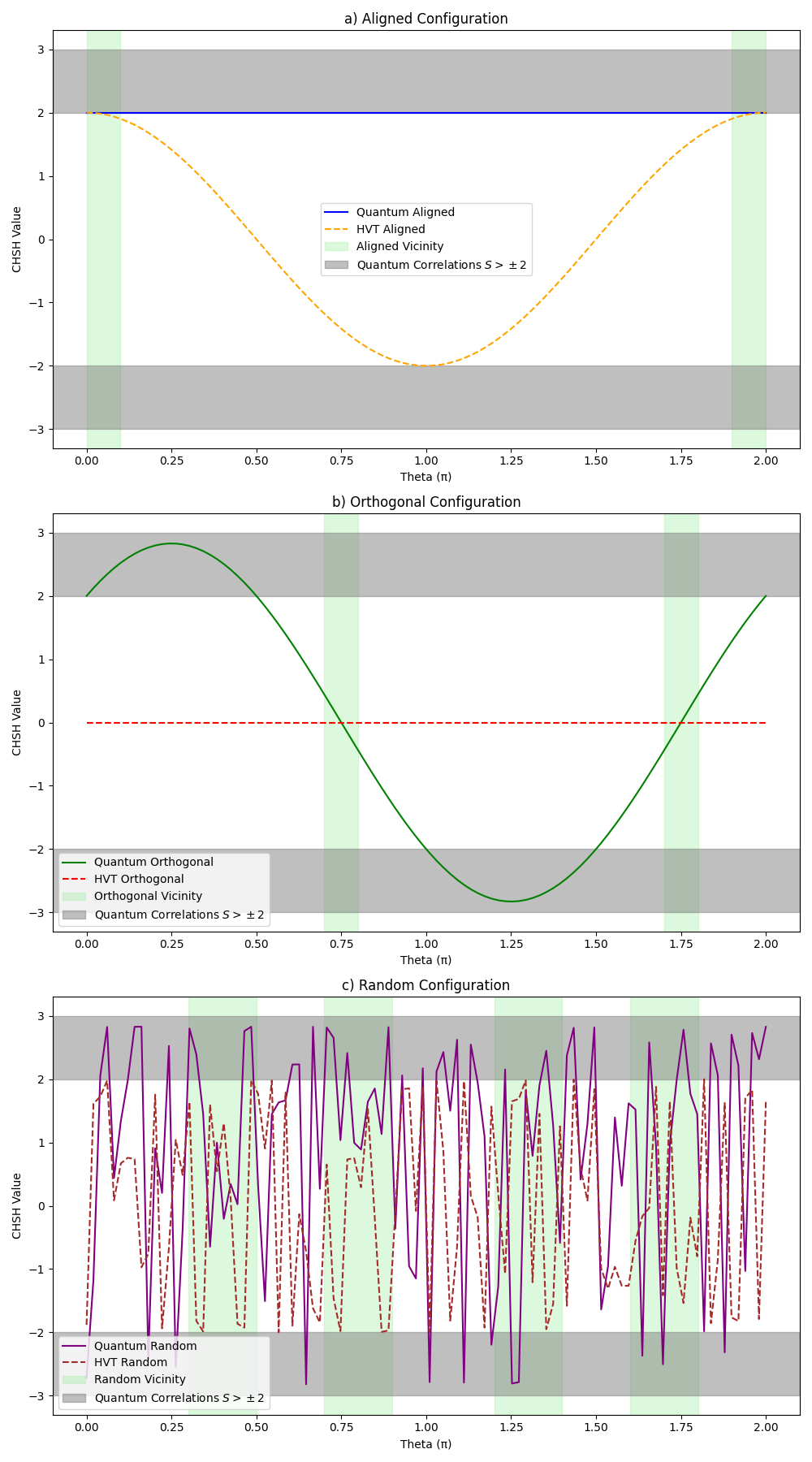}
    \caption{
    Simulation results comparing Quantum Mechanical (QM) predictions with Hidden Variable Theory (HVT) predictions across three different measurement configurations: Aligned, Orthogonal, and Random. The CHSH (Clauser-Horne-Shimony-Holt) \cite{clauser_proposed_1969} values are plotted against the measurement angle \( \theta \) (in radians). The shaded gray regions indicate the vicinities where quantum correlations exceed the classical threshold ( \( S > \pm 2 \) ), which signals a violation of the CHSH inequality, a hallmark of quantum entanglement. The light green regions denote the vicinities where the predictions from QM and HVT overlap, representing potential security vulnerabilities in quantum cryptographic protocols.
    }
    \label{fig:chsh_comparison}
\end{figure}

Our analysis focuses on multipartite quantum systems involving multiple entangled particles, which serve as a more realistic model for practical quantum cryptography. We identify critical points of convergence between BIs and HVTs, particularly where quantum mechanical predictions align with those from local hidden variable models \cite{aspect_experimental_1982}. When considered alongside the HUP, these convergence points expose previously unrecognized vulnerabilities in QKD protocols (See Figure \ref{fig:chsh_comparison}).

\subsubsection{Key Measurement Configurations and Their Security Implications}
It investigated three distinct measurement configurations (See Figures \ref{fig:measurement_configurations} and \ref{fig:chsh_comparison}): aligned, orthogonal, and random measurement directions. All measurement vectors are pointed in the aligned configuration in the same direction, maximizing correlation. The orthogonal configuration involves measurement vectors that are mutually perpendicular, leading to minimal or no correlation. The random configuration features measurement vectors oriented in arbitrary directions, resulting in varying levels of correlation. The analysis highlights how the correlation functions in these configurations reveal overlap between quantum mechanics and HVTs, particularly within vicinities defined using the Heisenberg Uncertainty Principle (HUP). The critical insight is that in these vicinities, where predictions from quantum mechanics and HVTs become indistinguishable, quantum cryptographic protocols could be vulnerable to exploitation.

The correlation functions are summarized in Table \ref{tab:summary_correlation}, which compares quantum mechanical predictions and those of HVTs across different measurement configurations. 

\begin{table}
    \centering
\caption{Summary of correlation functions for different measurement configurations in quantum systems, comparing predictions from Quantum Mechanics (QM) and Hidden Variable Theories (HVTs), and extending the analysis to their vicinities defined by the Heisenberg Uncertainty Principle (HUP). The table illustrates the correlation function values under Aligned, Orthogonal, and Random configurations, including their behavior in the vicinity of the HUP. Symbols used: \(E(\mathbf{a}_i, \mathbf{b}_j)\) represents the correlation function, where \(\mathbf{a}_i\) and \(\mathbf{b}_j\) are the measurement settings for the respective particles. (See \ref{derv:correlation}, \ref{deriv:chsh} and Annex \ref{annex:derivations} for details)}
    \label{tab:summary_correlation}
    \begin{tabular}{|c|c|c|}
        \hline
        \textbf{Measurement Configuration} & \textbf{QM (Bell) Prediction} & \textbf{HVTs Prediction} \\ \hline
        
        Aligned & \( E(\mathbf{a}_1, \mathbf{a}_1, \ldots, \mathbf{a}_1) \) & \( E(\mathbf{a}_1, \mathbf{a}_1, \ldots, \mathbf{a}_1) \) \\
        (\(\mathbf{a}_1 = \mathbf{a}_2 = \cdots = \mathbf{a}_n\)) & \( = \pm 1 \) & \( = \pm 1 \) \\ \hline
        
        Aligned Vicinity & \( E(\mathbf{a}) = E_{\text{HVT}}(\mathbf{a}) \) & \( E(\mathbf{a}) = E_{\text{HVT}}(\mathbf{a}) \) \\
        (\(\mathbf{a} \in [\mathbf{a}_0 - \Delta\mathbf{a}, \mathbf{a}_0 + \Delta\mathbf{a}]\)) & \( \approx \pm 1 \) & \( \approx \pm 1 \) \\ \hline
        
        Orthogonal & \( E(\mathbf{a}_1, \mathbf{a}_2, \ldots, \mathbf{a}_n) = 0 \) & \( E(\mathbf{a}_1, \mathbf{a}_2, \ldots, \mathbf{a}_n) = 0 \) \\
        (\(\mathbf{a}_i \perp \mathbf{a}_j\)) & & \\ \hline
        
        Orthogonal Vicinity & \( E(\mathbf{a}, \mathbf{b}) = E_{\text{HVT}}(\mathbf{a}, \mathbf{b}) \) & \( E(\mathbf{a}, \mathbf{b}) = E_{\text{HVT}}(\mathbf{a}, \mathbf{b}) \) \\
       \((\mathbf{a}, \mathbf{b}) \in\) & \( \approx  0 \) & \( \approx  0 \) \\
        \( [(\mathbf{a}_0, \mathbf{b}_0) - \Delta\theta, (\mathbf{a}_0, \mathbf{b}_0) + \Delta\theta]\) & & \\ \hline
        
        Random & \( \langle E(\mathbf{a}_1, \mathbf{a}_2, \ldots, \mathbf{a}_n) \rangle = 0 \) & \( \langle E_{\text{HVT}}(\mathbf{a}_1, \mathbf{a}_2, \ldots, \mathbf{a}_n) \rangle = 0 \) \\ \hline
        
        Random Vicinity & \( \langle E(\mathbf{a}, \mathbf{b}) \rangle = \langle E_{\text{HVT}}(\mathbf{a}, \mathbf{b}) \rangle \) & \( \langle E(\mathbf{a}, \mathbf{b}) \rangle = \langle E_{\text{HVT}}(\mathbf{a}, \mathbf{b}) \rangle \) \\
        (\(\mathbf{a}, \mathbf{b} \in \text{HUP Vicinity}\)) & \( \approx  0 \) & \( \approx  0 \) \\ \hline
    \end{tabular}
\end{table}

In Table \ref{tab:summary_chsh}, we have summarized the Clauser-Horne-Shimony-Holt (CHSH, an implementation of the Bell Inequality) parameter \( S \) values predicted by Quantum Mechanics (QM) and Hidden Variable Theories (HVTs) across three measurement configurations: aligned, orthogonal, and random. In the aligned configuration, QM predicts a maximal violation of the CHSH inequality with \( S = 2\sqrt{2} \), while HVTs limit \( S \) to 2. The orthogonal configuration yields no violation (\( S = 0 \)) for both QM and HVTs. For random configurations, QM still indicates an average violation (\( S \approx 2\sqrt{2} \)), contrasting with the classical bounds of HVTs (\( S \leq 2 \)).

\begin{table}
    \centering
\caption{Comparison of the Clauser-Horne-Shimony-Holt (CHSH) inequality parameter \( S \) values predicted by Quantum Mechanics (QM) and Hidden Variable Theories (HVTs) under different measurement configurations: aligned, orthogonal, and random. The table highlights how quantum mechanics can violate the CHSH inequality, contrasting this with the classical bounds imposed by HVTs.}
    \label{tab:summary_chsh}
    \begin{tabular}{|c|c|c|}
        \hline
        \textbf{Measurement Configuration} & \textbf{QM Prediction for \( S \)} & \textbf{HVTs Prediction for \( S \)} \\ \hline
        
        Aligned (\(\mathbf{a} = \mathbf{b}\)) & \( S = 2\sqrt{2} \) & \( S = 2 \) \\ \hline
        
        Orthogonal (\(\mathbf{a} \perp \mathbf{b}\)) & \( S = 0 \) & \( S = 0 \) \\ \hline
        
        Random & \( \langle S \rangle \) \( \approx 2\sqrt{2} \) & \( \langle S_{\text{HVT}} \rangle \) \( \leq 2 \)  \\ \hline
    \end{tabular}
\end{table}

\subsection{Practical Implications for QKD}
Our findings highlight a significant security risk in some QKD protocols when Alice and Bob measure entangled particles along the same axis. In such scenarios, the convergence of predictions from quantum mechanics and HVTs suggests that an adversary could exploit this alignment to perform undetected attacks, thereby compromising the security of the QKD protocol. This risk necessitates the careful selection of measurement settings to avoid these vulnerabilities.

\begin{figure}
    \centering
    \includegraphics[width=1\textwidth]{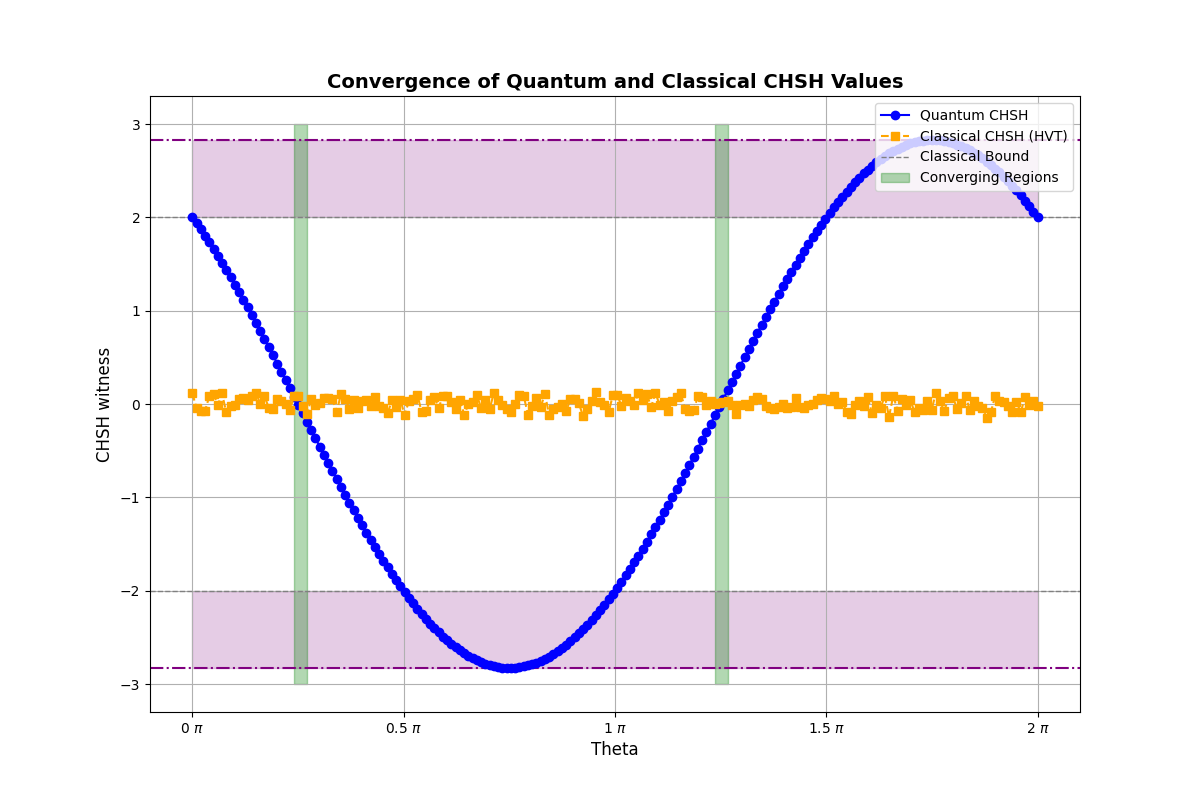}
    \caption{Results from numerical simulations demonstrating the convergence of Bell Inequalities and Hidden Variable Theories within specific measurement vicinities. The plot shows the CHSH (Clauser-Horne-Shimony-Holt) inequality witness values as a function of the measurement angle, illustrating regions where quantum mechanical predictions (blue curve) align with classical Hidden Variable Theories (orange curve). The shaded green regions highlight the vicinities where the quantum and classical descriptions converge, validating the theoretical predictions about potential vulnerabilities in existing quantum cryptographic protocols. The bounds for classical and quantum correlations are also indicated, providing a clear comparison between the two frameworks.}
    \label{fig:numerical_simulation_results}
\end{figure}

\begin{figure}
    \centering
    \includegraphics[width=1\textwidth]{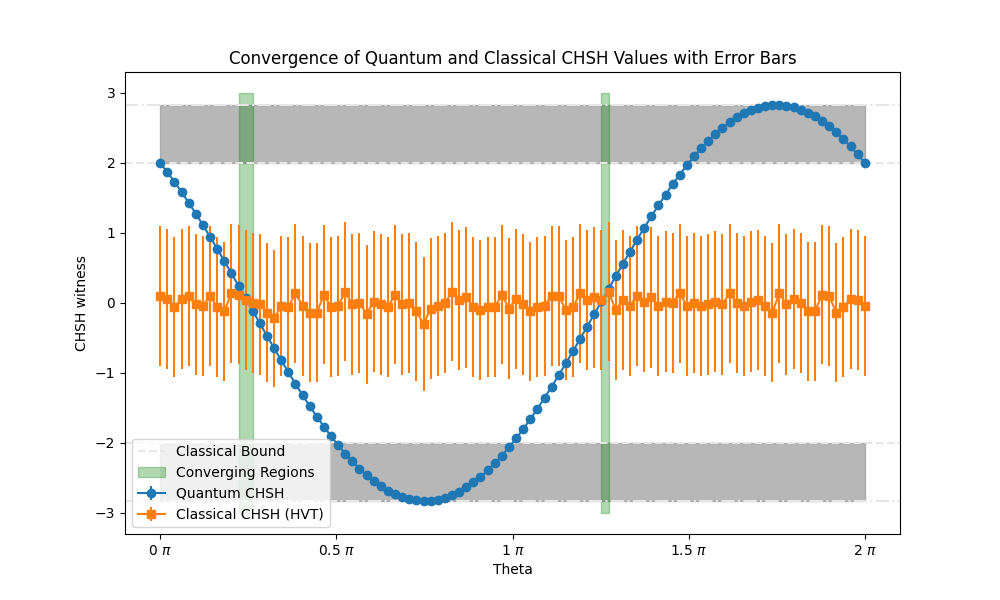}
    \caption{Convergence of Quantum and Classical CHSH (Clauser-Horne-Shimony-Holt) Values with Error Bars. The plot shows the CHSH witness values as a function of the measurement angle. Error bars indicate the statistical variability from multiple runs, providing a measure of the robustness of the simulation results. The green-shaded regions highlight the areas of convergence between quantum mechanical predictions (blue curve) and classical Hidden Variable Theories (orange curve).}
    \label{fig:CHSH_with_error_bars}
\end{figure}

The numerical simulations presented in Figures \ref{fig:numerical_simulation_results} and  \ref{fig:CHSH_with_error_bars} further illustrate the conditions under which quantum and classical descriptions converge, providing critical insights into potential vulnerabilities in quantum cryptographic protocols. These findings emphasize the importance of considering measurement configurations in maintaining the security of quantum systems.

\subsection{Practical Implications for Quantum Integrity}
The  \( S \) value, derived from the Clauser-Horne-Shimony-Holt (CHSH) inequality, is widely recognized in the quantum industry as a critical indicator of the strength of entanglement. It serves as a measure of quantum correlations, with values exceeding the classical bound of 2 suggesting a quantum advantage. In theoretical models and simulations, achieving an  \( S \) value greater than 2 is typically straightforward when using optimal measurement angles, as the CHSH inequality is inherently violated by quantum mechanics in these scenarios.

\begin{figure}
    \centering
    \includegraphics[width=1\textwidth]{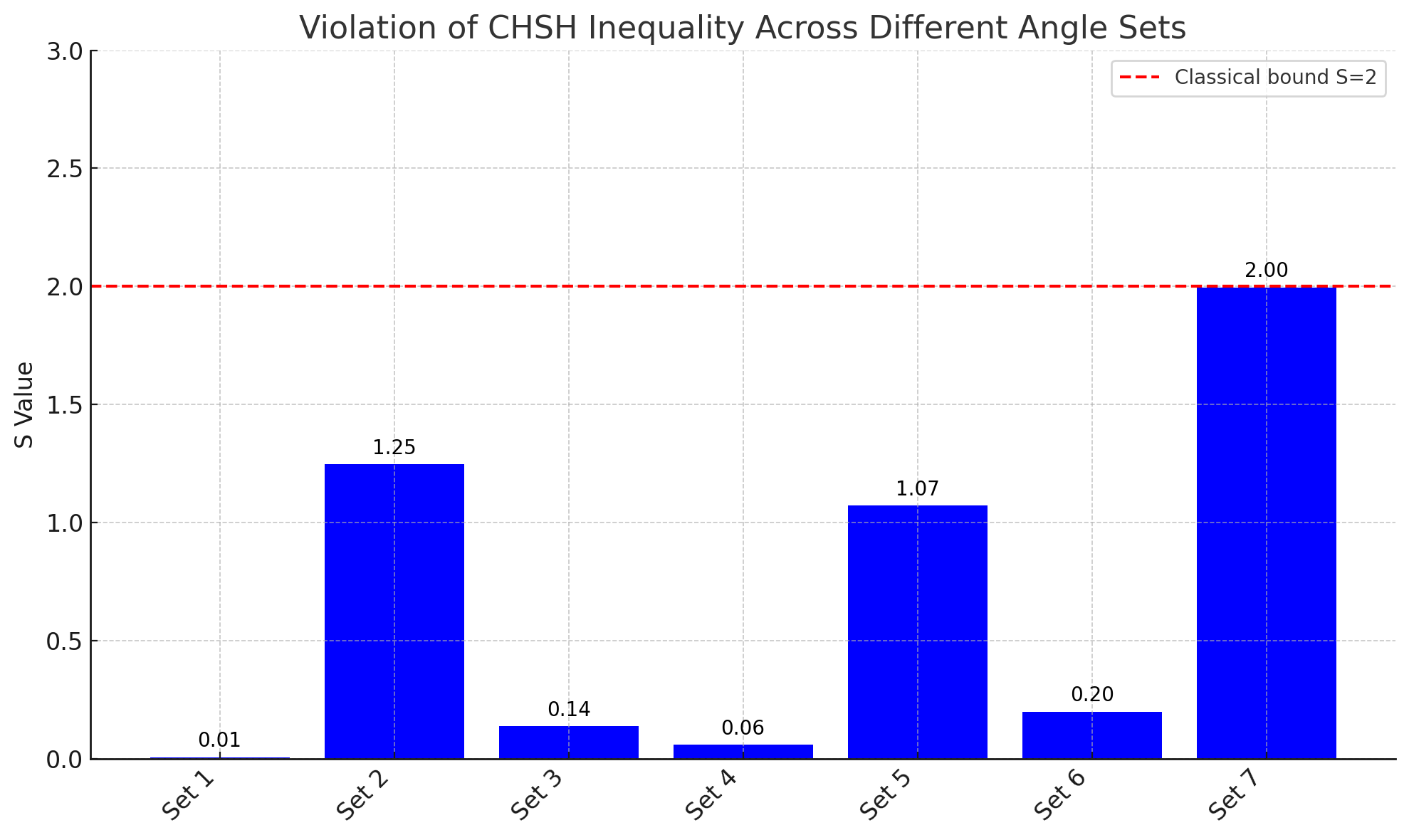}
    \caption{CHSH Inequality Across Different Angle Sets (Quantum Computer). The figure illustrates the results from an IBM quantum computer \cite{cloud_ibm_2024, quantum_ibm_2024}, showing the \( S \) values for various angle sets. Where none of the sets indicate a violation of the CHSH inequality, even when using maximal entangle angles, highlighting the impact of hardware noise and imperfections on quantum measurements. (See Annex \ref{annex:experiments:quantum})}
    \label{fig:CHSH_Quantum_Results}
\end{figure}

However, our experiments reveal a more nuanced picture when moving from theoretical and simulated environments to actual quantum hardware. As shown in Figure \ref{fig:CHSH_Quantum_Results}, even with optimal angles, reaching an \( S \) value that definitively indicates quantum entanglement can be challenging due to noise, decoherence, and other imperfections inherent in quantum devices. These factors often lead to \( S \) values that hover around or below the classical threshold, raising concerns about the reliability of \( S \) as a sole indicator of entanglement in practical applications.

These findings suggest a need for caution when relying on the \( S \) value in scenarios where quantum integrity is critical. In particular, systems that use \( S \) as a benchmark for validating quantum computations or for ensuring that a quantum system is operating as expected may be vulnerable. The difficulty in consistently achieving \( S \) values greater than 2 on real quantum hardware could lead to false positives, where a system is mistakenly believed to be functioning correctly or securely when it is not.

Moreover, our study highlights a potentially exploitable vulnerability within this framework. By leveraging the convergence of Bell Inequalities (BIs) and Hidden Variable Theories (HVTs), particularly within the vicinities where quantum and classical predictions overlap, an attacker could theoretically manipulate the system to produce misleading S values. This manipulation could deceive security protocols that depend on S as a verification tool, potentially allowing tampering with quantum computations or the simulation of quantum systems by classical means.

\begin{figure}
    \centering
    \includegraphics[width=1\textwidth]{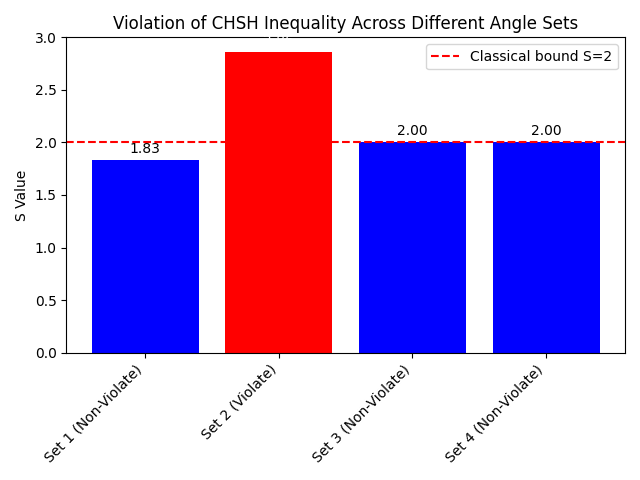}
    \caption{Violation of CHSH Inequality Across Different Angle Sets (Simulator). The figure shows the S values for different angle sets tested in the Qiskit AerSimulator \cite{aleksandrowicz_qiskit_2019, noauthor_qiskitqiskit-aer_2024}. The red bars indicate violations of the CHSH inequality where \( S > 2 \), confirming quantum correlations beyond classical limits. (See Annex \ref{annex:experiments:simulator})}
    \label{fig:CHSH_Simulator_Results}
\end{figure}

Figures \ref{fig:CHSH_Simulator_Results} and \ref{fig:CHSH_Quantum_Results} contrasts the results from a quantum simulator with those obtained from actual quantum hardware, emphasizing the discrepancy between theoretical predictions and practical outcomes. While the simulator indicates clear violations of the CHSH inequality, the quantum hardware struggles to reproduce these results consistently, underscoring the risks of over-reliance on \( S \) values in practical applications.

While the \( S \) value is a critical indicator in theoretical and simulated studies of quantum entanglement, our findings reveal that its practical application, especially in quantum integrity verification, requires more nuanced consideration. The observed limitations in our experiments underscore the potential for these tools to be exploited in real-world scenarios, where noise, decoherence, and other imperfections can obscure quantum advantages. Moreover, the convergence points between BIs and HVTs identified in our study suggest that reliance on the \( S \) value alone could expose quantum systems to new, sophisticated attack vectors. Therefore, additional metrics or complementary methods are crucial for robustly assessing the integrity and security of quantum systems under practical conditions, particularly as these vulnerabilities could be weaponized to undermine quantum protocols.

\section{Methods}\label{methods}
We explore the theoretical underpinnings and vulnerabilities of multipartite discrete variable quantum systems, focusing on the convergence of BIs and HVTs. Our analysis is grounded in 3 conditions: ideallized, simulated, and experimental. The idealized conditions allow us to isolate intrinsic quantum properties from practical implementation challenges. By examining the quantum system model, measurement settings, and hidden variable models under these ideal conditions, we identify potential points of convergence between quantum and classical descriptions. Furthermore, we extend our analysis by introducing the new concept of measurement vicinities, defined through the HUP, to assess the security implications of these convergence points. Numerical simulations are then employed to validate our theoretical predictions and explore their practical impact on the security of quantum protocols.  Finally, we test our simulations against a real physical quantum computer.
\subsection{Quantum System Model and Assumptions}
In this study, we investigate a multipartite discrete variable quantum system comprising \( n \) entangled particles. The analysis is conducted under idealized conditions to isolate theoretical vulnerabilities from practical implementation issues. These ideal conditions include perfect detectors, no environmental noise, and the absence of losses, ensuring that any observed effects are intrinsic to the quantum model rather than artifacts of experimental imperfections.

\subsubsection{System Configuration}
\begin{figure}
    \centering
    \includegraphics[width=1\textwidth]{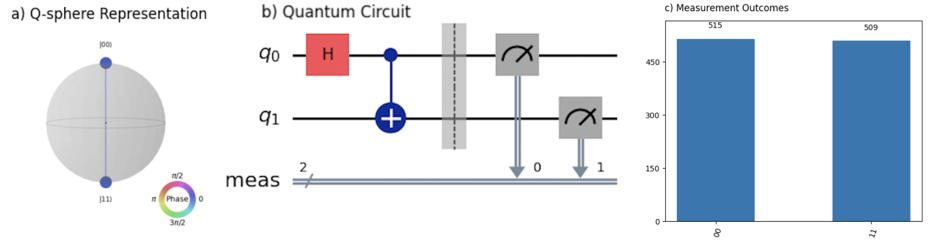}
    \caption{Visualization of the quantum system under study. The figure includes (a) the \textbf{Q-sphere} representation of the quantum state, (b) the \textbf{quantum circuit} used to generate a \textbf{Bell state}, and (c) the \textbf{measurement outcomes}. The Q-sphere represents the quantum state after entanglement, showing the probability amplitudes of the computational basis states. The quantum circuit illustrates the steps to generate and measure the Bell state, while the histogram of measurement outcomes verifies the entanglement by showing the expected equal probabilities for the $\mathbf{\ket{00}}$ and $\mathbf{\ket{11}}$ states.}
    \label{fig:quantum_system_schematic}
\end{figure}

The quantum system under consideration comprises \( n \) maximally entangled particles. Each particle is assumed to be part of an entangled state that can be described by a pure quantum state vector in a Hilbert space. The particles are distributed to different spatial locations, allowing for independent measurements of each particle. The entanglement between the particles ensures that measurement outcomes are correlated in a way that violates local realism, which is central to exploring BIs and HVTs. (See Figure \ref{fig:quantum_system_schematic} for a two particles system example)

\subsubsection{Measurement Settings}
Each particle in the system is subject to measurement along a chosen axis. The measurement settings for each particle are assumed to be freely chosen by the experimenters and are not influenced by any hidden variables or prior states. This assumption is crucial to maintain the independence required to test the validity of Bell Inequalities. The measurements are idealized as projective measurements, where the outcomes correspond to the eigenstates of the chosen measurement operators.

\subsubsection{Hidden Variable Models}
The study considers local hidden variable models, where the outcomes of quantum measurements are predetermined by hidden variables \( \lambda \). These variables are not accessible within the standard quantum mechanical framework but are assumed to exist under certain interpretations of quantum mechanics (e.g., EPR \cite{einstein_can_1935}). The hidden variable models employed here are deterministic, meaning that given the hidden variable \( \lambda \) and the measurement setting, the outcome is fully determined. This deterministic nature is contrasted with the probabilistic predictions of quantum mechanics, allowing for a thorough investigation of the convergence between BIs and HVTs.

\subsubsection{Ideal Conditions Assumptions}
To focus on the intrinsic properties of the quantum system, several ideal conditions are assumed:
\begin{itemize}
    \item \textbf{Perfect Detectors:} The detectors used in the measurement process are assumed to have 100\% efficiency, with no dead time, dark counts, or other imperfections. This assumption eliminates the possibility of false negatives or positives in the detection process.
    \item \textbf{No Environmental Noise:} The quantum system is considered completely isolated from its environment, preventing any decoherence or noise-induced errors. This isolation ensures the entanglement between the particles is preserved throughout the experiment.
    \item \textbf{Absence of Losses:} There are no photon losses, scattering, or absorption events in the system. All particles are assumed to reach their respective detectors without signal loss, ensuring that all entangled pairs are accounted for in the measurement process.
\end{itemize}

These ideal conditions provide a theoretical framework within which the fundamental properties and potential vulnerabilities of the quantum system can be explored without the confounding effects of practical imperfections.

\subsubsection{Relevance to Practical Implementations}
While the assumptions made in this study are idealized, they serve as a foundation for understanding the inherent risks in quantum systems. In practical implementations, factors such as detector inefficiency, noise, and losses would need to be considered, but the theoretical vulnerabilities identified here could manifest itself in real-world systems under less-than-ideal conditions. Future work must address these practical concerns and quantify how they might exacerbate or mitigate the risks identified in this study.

\subsection{Derivation of Correlation Functions}\label{derv:correlation}

\subsubsection{General Form of Correlation Functions}
To analyze the convergence of Bell Inequalities (BIs) \cite{bell_einstein_1964} and Hidden Variable Theories (HVTs) \cite{einstein_can_1935}, we derive the correlation functions for various measurement configurations, including aligned, orthogonal, and random measurement directions (see Annex \ref{annex:derivations} for full explanation ). The general form of the correlation function for \( n \) particles measured along directions \( \mathbf{a}_1, \mathbf{a}_2, \ldots, \mathbf{a}_n \) is expressed as:

\begin{equation}
\label{eq:deriv_v1}
E(\mathbf{a}_1, \mathbf{a}_2, \ldots, \mathbf{a}_n) = \int \rho(\lambda) \prod_{i=1}^{n} A_i(\mathbf{a}_i, \lambda) \, d\lambda,
\end{equation}

where \( \rho(\lambda) \) represents the probability distribution of the hidden variable \( \lambda \), and \( A_i(\mathbf{a}_i, \lambda) \) denotes the deterministic outcome of the measurement on the \( i \)-th particle.

For \textbf{aligned measurement directions}, where \( \mathbf{a}_1 = \mathbf{a}_2 = \cdots = \mathbf{a}_n = \mathbf{a} \), the correlation function simplifies to:

\begin{equation}
\label{eq:deriv_v2}
E(\mathbf{a}, \mathbf{a}, \ldots, \mathbf{a}) = \int \rho(\lambda) \left( \prod_{i=1}^{n} A_i(\mathbf{a}, \lambda) \right) \, d\lambda.
\end{equation}

In the case of \textbf{orthogonal measurement directions}, where each pair of directions is orthogonal (i.e., \( \mathbf{a}_i \perp \mathbf{a}_j \) for all \( i \neq j \)), the correlation function is expected to be zero under both quantum mechanics and HVTs:

\begin{equation}
\label{eq:deriv_v3}
E(\mathbf{a}_1, \mathbf{a}_2, \ldots, \mathbf{a}_n) = 0.
\end{equation}

For \textbf{random measurement directions}, we consider the average correlation function, which is given by:

\begin{equation}
\label{eq:deriv_v4}
\langle E(\mathbf{a}_1, \mathbf{a}_2, \ldots, \mathbf{a}_n) \rangle = \int \rho(\lambda) \left\langle \prod_{i=1}^{n} A_i(\mathbf{a}_i, \lambda) \right\rangle \, d\lambda.
\end{equation}

\subsubsection{Derivation of Correlation Functions for CHSH}\label{deriv:chsh}
In the context of the CHSH inequality \cite{clauser_proposed_1969}, we specifically focus on the correlation functions for measurements along two directions for two particles, typically labeled \( \mathbf{a}, \mathbf{a'} \) and \( \mathbf{b}, \mathbf{b'} \). The CHSH parameter \( S \) is computed using the following combination of correlation functions:

\begin{equation}
\label{eq:deriv_chsh1}
S = E(\mathbf{a}, \mathbf{b}) - E(\mathbf{a}, \mathbf{b'}) + E(\mathbf{a'}, \mathbf{b}) + E(\mathbf{a'}, \mathbf{b'}),
\end{equation}

where the correlation function \( E(\mathbf{a}, \mathbf{b}) \) is derived as:

\begin{equation}
\label{eq:deriv_chsh2}
E(\mathbf{a}, \mathbf{b}) = \int \rho(\lambda) A(\mathbf{a}, \lambda) B(\mathbf{b}, \lambda) \, d\lambda,
\end{equation}

and similarly for the other terms in \( S \). The value of \( S \) under quantum mechanics can exceed the classical limit of 2, reaching up to \( 2\sqrt{2} \) in the case of maximal violation. This is in contrast to the predictions of HVTs, where \( |S| \leq 2 \) must always hold.

\paragraph{Aligned Measurement Directions}
In the case where the measurement settings \( \mathbf{a} \) and \( \mathbf{b} \) are aligned with each other (i.e., \( \mathbf{a} = \mathbf{b} \) and \( \mathbf{a'} = \mathbf{b'} \)), the CHSH parameter \( S \) simplifies. Quantum mechanically, the expectation value for \( S \) in this scenario can reach the maximum quantum violation of \( 2\sqrt{2} \). This occurs when the angles between the measurement directions are optimized, specifically when the angle between \( \mathbf{a} \) and \( \mathbf{a'} \), as well as \( \mathbf{b} \) and \( \mathbf{b'} \), is \( \pi/2 \). For Hidden Variable Theories (HVTs), however, the maximum value of \( S \) in aligned directions remains limited to 2, in accordance with the CHSH inequality.

\paragraph{Orthogonal Measurement Directions}
For orthogonal measurement directions, where \( \mathbf{a} \) is orthogonal to \( \mathbf{b'} \) and \( \mathbf{a'} \) is orthogonal to \( \mathbf{b} \), the expectation value of \( S \) is zero, under both Quantum Mechanics and HVTs. The correlation functions \( E(\mathbf{a}, \mathbf{b'}) \) and \( E(\mathbf{a'}, \mathbf{b}) \) will yield zero due to the orthogonality of the measurement directions. As a result, no violation of the CHSH inequality is expected, even in quantum systems.

\paragraph{Random Measurement Directions}
When the measurement directions are chosen randomly, the CHSH parameter \( S \) must be averaged over many possible settings. Quantum mechanically, the average value of \( S \) can still exhibit a violation of the CHSH inequality, though the exact value of \( S \) will vary depending on the specific random choices of directions. The expected quantum value of \( S \) in such random configurations can approach \( 2\sqrt{2} \) on average, highlighting the robustness of quantum entanglement against the randomness of measurement settings. Conversely, HVTs will consistently yield \( |S| \leq 2 \) for random measurement directions, as they cannot surpass the local hidden variable bound.

See Annex \ref{annex:derivations} for more details.

\subsection{Numerical Simulations and Analysis of the particular CHSH Correlation Functions}
Numerical simulations for a two-particle system were conducted in this research to validate the theoretical predictions (See Annex \ref{annex:derivations}) regarding the convergence of BIs and HVTs within specified measurement vicinities. These simulations were performed under various initial conditions (See Annex \ref{annex:experiments}) and measurement settings to explore the robustness of these predictions across different quantum systems. To accomplish this, the CHSH inequality was simulated using IBM’s Quantum Cloud Platform (See Annex \ref{annex:experiments:quantum}) and IBM Qiskit Framework (AERSimulator \cite{aleksandrowicz_qiskit_2019, noauthor_qiskitqiskit-aer_2024}) (See Annex \ref{annex:experiments:simulator}) \cite{nielsen_quantum_2012}, with the analysis conducted via PyCharm.

The results of these simulations are presented in Figures \ref{fig:chsh_comparison}, \ref{fig:numerical_simulation_results}, \ref{fig:CHSH_with_error_bars}, \ref{fig:CHSH_Quantum_Results}, \ref{fig:CHSH_Simulator_Results} where they are compared against the theoretical predictions.

\section{Discussion}\label{discussion}

This study reveals significant theoretical vulnerabilities in quantum systems that rely on Bell Inequalities (BIs) for security, particularly within Quantum Key Distribution (QKD) protocols. These vulnerabilities are not merely extensions of implementation loopholes previously documented \cite{hensen_loophole-free_2015, xu_secure_2020, acin_device-independent_2007}. They arise from a fundamental convergence between Bell Inequalities (BIs) and Hidden Variable Theories (HVTs) under specific conditions. This convergence challenges the assumption that quantum mechanics inherently offers unassailable security through the violation of BIs.

Our analysis demonstrates that in certain measurement configurations, particularly those involving aligned measurement directions, quantum mechanical predictions may align with those from local HVTs. This alignment poses a critical security risk, as it undermines the expected distinction between quantum and classical predictions. Adversaries could potentially exploit these scenarios and compromise the integrity of QKD protocols.

\subsection{Implications for Quantum Cryptography}

The findings presented in this study underscore a substantial threat to the foundational security assumptions of quantum cryptographic protocols. The convergence of BIs and HVTs in specific scenarios suggests that QKD systems, which depend on the unique properties of quantum correlations, may be more susceptible to undetected attacks than previously recognized, particularly in situations where measurement settings might unintentionally align with those predicted by HVTs.

\subsection{Future Directions in Quantum Security Research}
Given the theoretical vulnerabilities identified, it is imperative to pursue experimental validation to assess the practical impact of these risks in operational quantum systems. Such experimental efforts will be crucial in determining how these vulnerabilities manifest in real-world scenarios and will provide valuable insights for developing more robust quantum protocols.

Furthermore, this study highlights the limitations of relying exclusively on BIs as the cornerstone of quantum security. Future research should explore integrating additional quantum mechanical principles or creating novel cryptographic frameworks designed to mitigate the risks associated with the convergence of quantum and classical predictions. Strengthening the resilience of quantum communication technologies will require a comprehensive approach that addresses these newly recognized threats and expands the security model beyond current paradigms.

\subsection{Conclusion and Outlook}
This study challenges the prevailing assumptions regarding the security of quantum systems, particularly those that rely on Bell Inequalities (BIs) as a cornerstone of their cryptographic protocols. The convergence points identified between quantum and classical theories reveal potential vulnerabilities that may compromise the integrity of quantum cryptographic systems. These findings suggest that the presumed invulnerability of quantum systems warrants closer scrutiny.

As the field of quantum cryptography progresses, the research community must rigorously examine these vulnerabilities. Developing more secure quantum communication technologies will depend on our ability to identify and mitigate these risks. We advocate for a collaborative effort among researchers to prioritize the design of resilient protocols for these newly uncovered threats and explore broader security frameworks.

While this study provides valuable insights into theoretical vulnerabilities in quantum cryptographic systems, it is important to acknowledge the limitations inherent in our approach. The findings primarily focus on idealized conditions, and further experimental validation is necessary to understand how these vulnerabilities manifest in practical quantum systems. Additionally, our reliance on certain assumptions, such as perfect detector efficiency and the absence of environmental noise, highlights the need for caution when extrapolating these results to real-world implementations. Future research should aim to experimentally validate these theoretical predictions and explore how practical limitations might influence the security of quantum systems.

The implications of this work extend beyond the specific vulnerabilities discussed; they raise fundamental questions about the robustness of current quantum cryptographic models. We encourage further theoretical and experimental investigations to assess the potential impact of these vulnerabilities on Quantum Key Distribution (QKD) and quantum integrity in general. Through such collective inquiry, the field of quantum cryptography can evolve to address the challenges posed by these findings, ensuring that the security of quantum communications remains uncompromised as we move into an era of increasingly sophisticated quantum technologies.

\section{Declarations}
\subsection{Acknowledgements}
This study was funded by \orgname{LAKES Environmental Research Inc.}, \orgaddress{\country{Canada}}.
\subsection{Conflict of interest}
All authors declare no conflict of interest. 
\subsection{Ethics approval and consent to participate}
This study did not involve human participants, animal subjects, or any other elements that require ethical approval. As such, ethics approval and consent to participate are not applicable.
\subsection{Consent for publication}
All authors have reviewed and approved the manuscript for publication. The authors have obtained consent from \orgname{LAKES Environmental Research Inc.} to publish the findings presented in this manuscript.
\subsection{Data availability}
This study is theoretical and did not generate or analyze any datasets. Therefore, data availability is not applicable.

\subsection{Materials availability}
No physical materials were used or generated in this research. Thus, materials availability is not applicable.

\subsection{Code availability}
The code used to generate the figures in this study is available from the corresponding author upon reasonable request.

\subsection{Author contribution}
J.R.R.B. was the lead author responsible for the study's conceptualization, experimentation, analysis, interpretation, and manuscript writing. J.V.G.T. contributed to the study's methodology and experimentation design, assisted with analysis, and provided significant input on interpreting results. R.A.F. reviewed the manuscript and contributed to analyzing and interpreting the study’s findings. All authors read and approved the final manuscript.

\begin{appendices}
\appendix
\section{Derivations}\label{annex:derivations}
\subsection{General Form of BIs for \textit{n} Particles}
For the purpose of this paper, we adopt a generalized formulation of BIs suitable for a system of \(n\) particles, each with two possible measurement settings (denoted by \(A_i\) and \(A'_i\) for the \(i\)th particle). This formulation accounts for the foundational work of Bell \cite{bell_einstein_1964} and Clauser et al. \cite{clauser_proposed_1969} and incorporates insights from more recent developments in the field \cite{brunner_bell_2014}. The general form of a Bell Inequality in this context can be expressed as:

\begin{equation}
\label{eq:a_deriv_1}
\left| \sum_{k} C_k \prod_{i=1}^{n} \langle A_i^{(k_i)} B_i^{(k_i)} \rangle \right| \leq B_{\text{local}},
\end{equation}

where:
\begin{itemize}
    \item \(C_k\) are coefficients that depend on the specific structure of the inequality and the number of particles involved.
    \item \(A_i^{(k_i)}\) and \(B_i^{(k_i)}\) represent the measurement outcomes for the \(i\)th particle, with \(k_i \in \{0, 1\}\) indicating the choice of measurement settings.
    \item \(B_{\text{local}}\) is the bound determined by local realism, which equals 2 for the simplest (e.g., CHSH-type) inequalities, but may vary in more complex scenarios.
\end{itemize}

\subsection{General Form of HVTs for \textit{n} Particles}
Hidden Variable Theories posit that the outcomes of quantum measurements are predetermined by hidden variables \( \lambda \), which are not accessible within standard quantum mechanics. For an \(n\)-particle system, the general form of the correlation function under HVTs is given by \cite{nielsen_quantum_2012, pearle_hidden-variable_1970}:

\begin{equation}
\label{eq:a_deriv_2}
E(\mathbf{a}_1, \mathbf{a}_2, \ldots, \mathbf{a}_n) = \int \rho(\lambda) \prod_{i=1}^{n} A_i(\mathbf{a}_i, \lambda) \, d\lambda,
\end{equation}
where:
\begin{itemize}
    \item \( \mathbf{a}_i \) represents the measurement setting for the \(i\)th particle.
    \item \( \lambda \) denotes the hidden variables with a probability distribution \( \rho(\lambda) \).
    \item \( A_i(\mathbf{a}_i, \lambda) \) are deterministic functions (taking values \( \pm 1 \)) that describe the measurement outcomes for each particle \(i\).
\end{itemize}

\subsection{Application to Electron Spin Angle Measurements}
Let \(\theta_0\) be a parameter (such as an angle) where the predictions of BIs and HVTs coincide \cite{heisenberg_uber_1927, horodecki_quantum_2009}:
\begin{equation}
\label{eq:a_deriv_3}
\mathcal{B}(\theta_0) = \mathcal{H}(\theta_0),
\end{equation}
where \(\mathcal{B}\) represents the result from the Bell inequality, and \(\mathcal{H}\) represents the corresponding prediction from the hidden variable theory.

To explore the vicinity around \(\theta_0\), we define a range of values \(\theta \in [\theta_0 - \Delta\theta, \theta_0 + \Delta\theta]\), where \(\Delta\theta\) is the uncertainty in the parameter \(\theta\), determined by the Heisenberg Uncertainty Principle:
\begin{equation}
\label{eq:a_deriv_4}
\Delta\theta \approx \frac{\hbar}{2\Delta J},
\end{equation}
where \(\Delta J\) represents the uncertainty in the conjugate variable to \(\theta\) (e.g., angular momentum if \(\theta\) is an angle).

This analysis suggests that within the range:
\begin{equation}
\label{eq:a_deriv_5}
\theta \in [\theta_0 - \frac{\hbar}{2\Delta J}, \theta_0 + \frac{\hbar}{2\Delta J}],
\end{equation}
The predictions of BIs and HVTs might remain indistinguishable, potentially exposing vulnerabilities in quantum systems if not properly accounted for.

\subsection{Derivation of the CHSH Inequality as a Particular Case of the General Form of BIs}

The general form of BIs for a system of \(n\) particles, each with two possible measurement settings (denoted by \(A_i\) and \(A'_i\) for the \(i\)th particle), can be expressed as \cite{clauser_proposed_1969, wallman_generating_2011}:

As described in equation \ref{eq:a_deriv_1}, to derive the CHSH inequality, consider the scenario where \(n = 2\), indicating the presence of two particles, each endowed with two distinct measurement settings. Designate the measurement settings for the first particle as \(A_1\) and \(A_1'\), and for the second particle, as \(B_2\) and \(B_2'\). Under these conditions, the general formulation of the Bell inequality reduces to:

\begin{equation}
\label{eq:a_deriv_7}
\left| C_1 \langle A_1 B_2 \rangle + C_2 \langle A_1 B_2' \rangle + C_3 \langle A_1' B_2 \rangle + C_4 \langle A_1' B_2' \rangle \right| \leq B_{\text{local}}.
\end{equation}

The CHSH inequality is obtained by choosing the coefficients \(C_k\) as follows: \(C_1 = 1\), \(C_2 = -1\), \(C_3 = 1\), and \(C_4 = 1\). Substituting these values into the general form gives:

\begin{equation}
\label{eq:a_deriv_8}
\left| \langle A_1 B_2 \rangle - \langle A_1 B_2' \rangle + \langle A_1' B_2 \rangle + \langle A_1' B_2' \rangle \right| \leq 2.
\end{equation}

This is the standard form of the CHSH inequality:

\begin{equation}
\label{eq:a_deriv_9}
S = |E(a,b) - E(a,b') + E(a',b) + E(a',b')| \leq 2,
\end{equation}

where \(E(a,b)\) denotes the expectation value of the measurement outcome product for settings \(a\) and \(b\) \cite{wallman_generating_2011}.

A violation, where \(S > 2\), suggests a quantum correlation that cannot be explained by any local hidden variable theory \cite{clauser_proposed_1969}.

Thus, the CHSH inequality is a specific case of the general Bell inequality for two particles with two measurement settings, illustrating the relationship between the general form of BIs and particular instances like the CHSH inequality.

\section{CHSH Inequality Experiments}\label{annex:experiments}

\subsection{Introduction to the Experiments}
In this annex, we present an analysis of the experiments conducted to test the Clauser-Horne-Shimony-Holt (CHSH) inequality using a quantum simulator (AerSimulator \cite{aleksandrowicz_qiskit_2019, noauthor_qiskitqiskit-aer_2024}) and an IBM quantum computer \cite{cloud_ibm_2024, quantum_ibm_2024}. The experiments aimed to compare theoretical predictions with empirical results, providing insights into the implementation of quantum protocols on real quantum hardware.

\subsection{Experimental Setup and Methodology}
The experiments were designed to evaluate the CHSH inequality by measuring the correlations between two entangled qubits. The core of the experimental setup involves creating a Bell state, followed by applying measurement operations in various bases defined by specific angles. The primary goal was to calculate the CHSH S value for different sets of measurement angles to observe whether quantum entanglement leads to a violation of the classical bound \( S \leq 2 \).

\subsubsection{Quantum Circuit Design}
The quantum circuit used in these experiments consists of the following key components:
\begin{enumerate}
    \item \textbf{Hadamard Gate (H):} The Hadamard gate is applied to the first qubit to create a superposition state. This operation is crucial for initiating the entanglement process in the subsequent step.
    \item \textbf{CNOT Gate:} The Controlled-NOT (CNOT) gate is applied with the first qubit as the control and the second qubit as the target. This gate entangles the two qubits, creating a Bell state, which is a maximally entangled quantum state.
    \item \textbf{Rotational Gates (RY):} Rotations around the Y-axis are applied to each qubit. These rotations allow measurements to be performed on bases other than the standard computational basis. The angles for these rotations are chosen according to the specific set being tested, influencing the correlation measurements.
    \item \textbf{Measurement:} The qubits are finally measured on a computational basis. The results of these measurements are used to calculate the expectation values required for computing the CHSH S value.
\end{enumerate}

\subsection{Simulator Experiment}\label{annex:experiments:simulator}
The first set of experiments was conducted using the Qiskit AerSimulator \cite{aleksandrowicz_qiskit_2019, noauthor_qiskitqiskit-aer_2024}, a high-performance simulator that allows for the execution of quantum circuits without the noise and decoherence present in real quantum hardware.

\subsubsection{Sets of Measurement Angles}
The following sets of angles were tested in the simulator (see Table \ref{tab:chsh_comparison} for a detailed comparison of theoretical and simulator values):

\begin{enumerate}
    \item \textbf{Set 1 (Non-Violate):} \((\frac{\pi}{6}, \frac{\pi}{6}), (\frac{\pi}{6}, \frac{2\pi}{6}), (\frac{2\pi}{6}, \frac{\pi}{6}), (\frac{2\pi}{6}, \frac{\pi}{4})\)
    \item \textbf{Set 2* (Violate):} \((0, \frac{\pi}{8}), (0, \frac{3\pi}{8}), (\frac{\pi}{8}, 0), (\frac{3\pi}{8}, \frac{\pi}{4})\)
    \item \textbf{Set 3 (Non-Violate):} \((0, 0), (\frac{\pi}{4}, \frac{\pi}{4}), (0, 0), (\frac{\pi}{4}, \frac{\pi}{4})\)
    \item \textbf{Set 4 (Non-Violate):} \((\frac{\pi}{8}, \frac{\pi}{8}), (\frac{3\pi}{8}, \frac{3\pi}{8}), (\frac{\pi}{8}, \frac{\pi}{8}), (\frac{3\pi}{8}, \frac{3\pi}{8})\)
\end{enumerate}

In Table \ref{tab:chsh_comparison}, Set 2 is marked with an asterisk (\(*\)) because it is theoretically expected to exhibit maximal entanglement, leading to a significant violation of the CHSH inequality, where \( S > 2 \).

\begin{table}
\centering
\caption{Comparison of Theoretical and Simulator Values for CHSH Inequality. The angle sets marked with an asterisk \((*)\) are those theoretically expected to exhibit maximal entanglement, leading to a violation of the CHSH inequality ( \(S > 2\)).}
\label{tab:chsh_comparison}
\begin{tabular}{|c|c|c|c|}
\hline
\textbf{Set} & \textbf{Angles} & \textbf{Theoretical S} & \textbf{Simulator S} \\ \hline
Set 1 & \(\left(\frac{\pi}{6}, \frac{\pi}{6}\right), \left(\frac{\pi}{6}, \frac{2\pi}{6}\right), \left(\frac{2\pi}{6}, \frac{\pi}{6}\right), \left(\frac{2\pi}{6}, \frac{\pi}{4}\right)\) & 1.8328 & 1.8328 \\ & & & \\ \hline
Set 2* & \(\left(0, \frac{\pi}{8}\right), \left(0, \frac{3\pi}{8}\right), \left(\frac{\pi}{8}, 0\right), \left(\frac{3\pi}{8}, \frac{\pi}{4}\right)\) & 2.8562 & 2.8562 \\ & & & \\  \hline
Set 3 & \(\left(0, 0\right), \left(\frac{\pi}{4}, \frac{\pi}{4}\right), \left(0, 0\right), \left(\frac{\pi}{4}, \frac{\pi}{4}\right)\) & 2.0000 & 2.0000 \\ & & & \\  \hline
Set 4 & \(\left(\frac{\pi}{8}, \frac{\pi}{8}\right), \left(\frac{3\pi}{8}, \frac{3\pi}{8}\right), \left(\frac{\pi}{8}, \frac{\pi}{8}\right), \left(\frac{3\pi}{8}, \frac{3\pi}{8}\right)\) & 2.0000 & 2.0000 \\ & & & \\  \hline
\end{tabular}
\end{table}

\subsubsection{Results and Discussion}
The results from the AerSimulator \cite{aleksandrowicz_qiskit_2019, noauthor_qiskitqiskit-aer_2024}, as illustrated in Figure \ref{fig:CHSH_Simulator_Results}, confirm that only Set 2 produced an S value that exceeded the classical bound of 2, achieving an S value of 2.8562. The other sets resulted in S values of 1.8328, 2.0000, and 2.0000, respectively. These outcomes validate the theoretical predictions, demonstrating that the quantum correlations present in the Bell state can indeed surpass classical limits under the right conditions, specifically with the angles used in Set 2.

\subsection{Quantum Computer Experiment}\label{annex:experiments:quantum}
The experiment was replicated on an IBM quantum computer \cite{cloud_ibm_2024, quantum_ibm_2024} using Qiskit \cite{aleksandrowicz_qiskit_2019} to validate the results. It is important to note that to evaluate maximum entanglement, we employed different sets of angles in the quantum computer experiments, as described below:

\subsubsection{Sets of Measurement Angles}
The following sets of angles were tested on the quantum computer:

\begin{enumerate}
    \item \textbf{Set 1:} \((0, \frac{\pi}{4}, \frac{\pi}{8}, \frac{3\pi}{8})\)
    \item \textbf{Set 2:} \((0, \frac{\pi}{8}, \frac{\pi}{4}, \frac{3\pi}{8})\)
    \item \textbf{Set 3:} \((0, \frac{\pi}{6}, \frac{\pi}{3}, \frac{\pi}{4})\)
    \item \textbf{Set 4:} \((0, \frac{\pi}{4}, \frac{\pi}{4}, \frac{\pi}{4})\)
    \item \textbf{Set 5:} \((0, \frac{\pi}{3}, \frac{\pi}{6}, \frac{\pi}{8})\)
    \item \textbf{Set 6:} \((0, \frac{\pi}{5}, \frac{\pi}{3}, \frac{\pi}{4})\)
    \item \textbf{Set 7:} \((\frac{\pi}{8}, \frac{\pi}{8}, \frac{\pi}{8}, \frac{\pi}{8})\)
    \item \textbf{Set 8:} \((\frac{\pi}{3}, \frac{\pi}{4}, \frac{\pi}{6}, \frac{\pi}{8})\)
\end{enumerate}

In Table \ref{tab:chsh_comparison_qc}, Sets 1 and 2 use angles that are theoretically expected to exhibit significant quantum correlations, but only Set 2 is aligned for a stronger CHSH violation (\(S > 2\)). Set 7 uses identical angles, which do not typically produce maximal entanglement, and thus the CHSH inequality is not strongly violated in this case.

\begin{table}
\centering
\caption{Comparison of Theoretical and Quantum Computer Values for CHSH Inequality. The angle sets marked with an asterisk \((*)\) are those theoretically expected to exhibit stronger quantum correlations, leading to a violation of the CHSH inequality ( \(S > 2\)).}
\label{tab:chsh_comparison_qc}
\begin{tabular}{|c|c|c|c|}
\hline
\textbf{Set} & \textbf{Angles} & \textbf{Theoretical S} & \textbf{Quantum Computer S} \\ \hline
Set 1 & \(\left(0, \frac{\pi}{4}\right), \left(\frac{\pi}{8}, \frac{3\pi}{8}\right)\) & 2.0000 & 0.0059 \\ & & & \\ \hline
Set 2* & \(\left(0, \frac{\pi}{8}\right), \left(\frac{\pi}{4}, \frac{3\pi}{8}\right)\) & 2.8284 & 1.2461 \\ & & & \\ \hline
Set 3 & \(\left(0, \frac{\pi}{6}\right), \left(\frac{\pi}{3}, \frac{\pi}{4}\right)\) & 2.0000 & 0.1367 \\ & & & \\ \hline
Set 4 & \(\left(0, \frac{\pi}{4}\right), \left(\frac{\pi}{4}, \frac{\pi}{4}\right)\) & 2.0000 & 0.0586 \\ & & & \\ \hline
Set 5 & \(\left(0, \frac{\pi}{3}\right), \left(\frac{\pi}{6}, \frac{\pi}{8}\right)\) & 2.0000 & 1.0723 \\ & & & \\ \hline
Set 6 & \(\left(0, \frac{\pi}{5}\right), \left(\frac{\pi}{3}, \frac{\pi}{4}\right)\) & 2.0000 & 0.1992 \\ & & & \\ \hline
Set 7 & \(\left(\frac{\pi}{8}, \frac{\pi}{8}\right), \left(\frac{\pi}{8}, \frac{\pi}{8}\right)\) & 2.0000 & 1.9961 \\ & & & \\ \hline
Set 8 & \(\left(\frac{\pi}{3}, \frac{\pi}{4}\right), \left(\frac{\pi}{6}, \frac{\pi}{8}\right)\) & 2.0000 & 1.0000 \\ & & & \\ \hline
\end{tabular}
\end{table}

\subsubsection{Results and Analysis}
The results from the quantum computer are shown in Figure \ref{fig:CHSH_Quantum_Results}. The quantum hardware provided results that consistently fell below the S value of 2, indicating no violation of the CHSH inequality in any of the tested sets. The highest S value observed was \(1.9961\) for Set 7, which, despite being close to the classical limit, did not exceed it. This particular set used identical angles, which are not expected to produce maximal entanglement, thus explaining the S value being below 2. 

These results highlight the impact of noise and imperfections in current quantum hardware, which can prevent the consistent realization of quantum violations that are predicted by theory and observed in ideal simulations. Theoretical predictions indicate that Set 2 should exhibit a significant violation of the CHSH inequality (with an expected S value close to \(2.8284\)); however, the actual quantum computer results only reached \(1.2461\), further emphasizing the gap between idealized quantum behavior and current experimental capabilities.

\subsection{Conclusion}
The experiments conducted using both the AerSimulator \cite{aleksandrowicz_qiskit_2019, noauthor_qiskitqiskit-aer_2024} and an IBM quantum computer \cite{cloud_ibm_2024, quantum_ibm_2024} illustrate the substantial gap between theoretical predictions and the current practical implementations of quantum entanglement as measured by the CHSH inequality. The simulator results demonstrated the expected violations of the CHSH inequality under ideal conditions, confirming the theoretical predictions with a clear violation in Set 2. However, the quantum hardware experiments consistently failed to achieve these violations, with all observed S values remaining below the critical threshold of 2.

The highest \(S\) value obtained on the quantum computer was 1.9961 in Set 7, which, while close to the classical limit, did not surpass it. The expected significant violation in Set 2, predicted by theory to yield an \(S\) value close to \(2.8284\), was not realized in practice, with the quantum computer only reaching an \(S\) value of \(1.2461\). These results underscore the challenges posed by noise, decoherence, and other imperfections in current quantum hardware.

\end{appendices}

\bibliography{sn-bibliography}

\end{document}